\documentclass[aps,12pt,showpacs]{revtex4}
\usepackage{amssymb}
\usepackage{graphicx}

\usepackage{amsmath}
\usepackage{bm}

\begin{document}

\title{Magnetosonic solitons in a Fermionic quantum plasma}
\author{M. Marklund, B. Eliasson, and P. K. Shukla}
\affiliation{Department of Physics, Ume{\aa} University, SE--901 87 Ume{\aa},
Sweden}
\received{10 April 2007}
\revised{17 September 2007}

\begin{abstract}
Starting from the governing equations for a quantum magnetoplasma including the quantum Bohm
potential and electron spin-$1/2$ effects, we show that the system of quantum magnetohydrodynamic
(QMHD) equations admit rarefactive solitons due to the balance between nonlinearities and
quantum diffraction/tunneling effects. It is found that the electron spin-$1/2$ effect introduces a
 pressure-like term with negative sign in the QMHD equations, which modifies the shape of the solitary
magnetosonic waves and makes them wider and shallower. Numerical simulations of the
time-dependent system shows the development of rarefactive QMHD solitary waves that are
modified by the spin effects.
\end{abstract}

\pacs{52.27.-h, 52.27.Gr, 67.57.Lm}

\maketitle

%%%%%%%%%%%%%%%%%%%
\section{Introduction}
%%%%%%%%%%%%%%%%%%%

There is currently a great deal of interest in collective quantum effects in plasmas \cite
{haas-etal1,anderson-etal,haas-etal2,haas,garcia-etal,marklund,shukla-stenflo,shukla,Shukla-Eliasson,
shukla-etal,manfredi,marklund-brodin}; many of these studies are motivated by recent
experimental progress and techniques \cite{marklund-shukla,exp1,exp2,glenzer-etal} and also by possible
astrophysical applications \cite{melrose,melrose-weise,baring-etal,harding-lai,marklund-brodin}.
In particular, magnetohydrodynamic (MHD) plasmas are of interest in such astrophysical
applications. However, in strong magnetic fields, single electron effects that depend on the
electron spin properties, such as Landau quantization, will be important. It
is thus not surprising that collective spin effects can influence the wave propagation
in a strongly magnetized quantum plasma \cite{marklund-brodin,cowley,kulsrud,brodin-marklund}.
Moreover, the recent progress in producing ultra-cold plasmas in terms of Rydberg states \cite{li-etal,fletcher-etal} may offer an
interesting experimental environment for quantum plasma dynamics. In such cold plasmas, the 
thermal energy of the particles can be very small compared to the Zeeman energy of the particles in 
an magnetic fields. Thus, collective spin properties of quantum plasmas may be possible to detect in a near future.

In this Brief Report, we will show that the balance between the nonlinear plasma and quantum effects
gives rise to magnetosonic solitons. Using the governing equations for QMHD plasmas with tunneling
and spin effects included, we derive a Sagdeev potential for the one-dimensional system.
We show that in a magnetized quantum plasma, the electron spin-$1/2$ effect can strongly
modify the amplitude and width of rarefactive solitons.

%%%%%%%%%%%%%%%%%%%
\section{Governing equations}
%%%%%%%%%%%%%%%%%%%

We begin by presenting the general governing equations for a quantum
magnetoplasma in which the electron$-1/2$ spin effect are included.
We define the total mass density $\rho \equiv (m_{e}n_{e}+m_{i}n_ {i})$,
the center-of-mass fluid flow velocity $\bm{V}\equiv (m_{e}n_{e}\bm{v}%
_{e}+m_{i}n_{i}\bm{v}_{i})/\rho $, and the current density $\bm{j}=- en_{e}%
\bm{v}_{e} + en_{i}\bm{v}_{i}$. Here $m_e$ ($m_i$) is the electron (ion) mass,
$n_e$ ($n_i$) is the electron (ion) number density, $\bm{v}_e$ ($\bm{v}_i$) is the
electron (ion) fluid velocity, and $e$ is the magnitude of the electron charge. From
the general set of spin-fluid equations \cite{marklund-brodin}
the corresponding QMHD equations can be derived \cite{brodin-marklund}. From these,
we immediately obtain the continuity equation

\begin{equation}
\frac{\partial \rho }{\partial t}+\bm{\nabla}\cdot (\rho \bm{V})=0 .
\label{eq:mhd-cont}
\end{equation}
Assuming the quasi-neutrality, i.e.\ $n_{e}\approx n_{i}$, the momentum conservation
equation reads

\begin{equation}
\rho \left( \frac{\partial }{\partial t}+\bm{V}\cdot \bm{\nabla} \right) %
\bm{V}=\bm{j}\times \bm{B}-\bm {\nabla}P+%
\bm{F}_{Q},  \label{eq:mhd-mom}
\end{equation}
where $P$ is the scalar pressure in the center-of-mass frame, the current is given by
$\bm{j} = \mu_0^{-1}\bm{\nabla}\times(\bm{B} - \mu_0\bm{M})$, $\bm{M} = (\mu_B\rho/m_i)\tanh( \mu_{B}B/k_{B}T_{e})\hat{\bm{B}}$ is the plasma magnetization due to the electron spin,  
and \cite{marklund-brodin,brodin-marklund}

\begin{equation}
  \bm{F}_{Q} = \frac{\hbar^2\rho}{2m_em_i}\bm{\nabla}\left( \frac{1}{\sqrt{\rho}}\nabla^{2}\sqrt{\rho}\right)
    + \frac{\mu_B\rho}{m_i}\tanh\left( \frac{\mu _{B}B}{k_{B}T_{e}}\right) \bm{\nabla} B
\label{eq:f-closed}
\end{equation}
is the quantum force due to collective tunneling and spin alignment. Here
$\mu_B = e\hbar/2m_e$ is the magnitude of the Bohr magneton, $\hbar$ is Planck constant divided by
$2\pi$, and $c$ is the speed of light in vacuum.  The  generalized Faraday law takes the form

\begin{equation}
\frac{\partial \bm{B}}{\partial t}=\bm{\nabla}\times \Bigg\{\bm{V}\times %
\bm{B}-\frac{\left[\bm{\nabla}\times (\bm{B} - \mu_0\bm{M})\right]\times \bm {B}}{%
en_e\mu _{0}}-\eta \bm{j}-\frac{m_{e}}{e^{2}\mu _{0}}\left[ \frac{\partial }{%
\partial t}-\left( \frac{\bm{\nabla}\times \bm{B}}{e\mu _{0}n_e}\right) \cdot %
\bm{\nabla}\right] \frac{\bm{\nabla}\times \bm{B}}{n_e}-\frac{\bm{F}_{Q}}{en_e}%
\Bigg\},
\label{eq:dynamo}
\end{equation}
where  $\eta$ is the plasma resistivity.

%%%%%%%%%%%%%%%%%%%%%%%%%
\section{Spin solitons}
%%%%%%%%%%%%%%%%%%%%%%%%%

Next, we assume that the magnetic field is along the
$z$ direction such that $\bm{B} = B(x,t)\hat{\bm{z}}$, while
we have the velocity $\bm{V} = V(x,t)\hat{\bm{x}}$ and the
density $\rho(x,t)$. With this, the governing
equations reduce to

\begin{equation}
  \frac{\partial\rho}{\partial t} + \frac{\partial}{\partial x}(\rho V) = 0 ,
\end{equation}

\begin{eqnarray}
  &&
  \frac{\partial V}{\partial t} + V\frac{\partial V}{\partial x} =
    -\frac{B}{\mu_0\rho}\frac{\partial B}{\partial x} - C_s^2\frac{\partial}{\partial x}\ln\rho
 \nonumber \\ &&\qquad\qquad
    + 2c^2\lambda_C^2\frac{m_e}{m_i}\frac{\partial}{\partial x}\left(
      \frac{1}{\sqrt{\rho}}\frac{\partial^2\sqrt{\rho}}{\partial x^2}
    \right)
    + \frac{\mu_B}{m_i\rho}\frac{\partial}{\partial x}\left[ 
      \rho B \tanh\left(
        \frac{\mu_BB}{k_BT_e}
      \right)
    \right] ,
\label{eq:mom}
\end{eqnarray}
and

\begin{equation}\label{eq:ohm}
  \frac{\partial B}{\partial t} + \frac{\partial}{\partial x}(BV) - \lambda\frac{\partial^2B}{\partial x^2} = 0 .
\end{equation}
Here $\lambda_C = c/\omega_C =\hbar/2m_ec$ is the Compton wavelength, $\omega_C$ is the
Compton frequency, $C_s = [k_B(T_e + T_i)/m_i]^{1/2}$ is the sound speed, $\lambda = \eta/\mu_0$ is the
magnetic diffusivity, the last term in Eq. (\ref{eq:mom}) is the spin force divided by $m_i$,
and we have neglected the inertial term in the Faraday law (\ref{eq:ohm}).

If the resistivity is weak, we may neglect the last term in the Faraday law (\ref{eq:ohm}), and obtain the
frozen-in-field condition $\rho = \rho_0b$,where $b = B/B_0$, with the background values denoted by the zero index.
Then, Eqs.\ (\ref{eq:mom}) and (\ref{eq:ohm}) form a closed system, taking the form

\begin{eqnarray}
  &&\!\!\!\!\!\!\!\!\!\!\!\!
  \frac{\partial V}{\partial t} + \frac{\partial}{\partial x}\left( \frac{V^2}{2} \right) =
    - C_A^2\frac{\partial b}{\partial x}
    - C_s^2\frac{\partial}{\partial x}\ln b
  \nonumber \\ &&\!\!\!\!\!\!\!\!
    + 2c^2\lambda_C^2\frac{m_e}{m_i}\frac{\partial}{\partial x}\left(
      \frac{1}{\sqrt{b}}\frac{\partial^2\sqrt{b}}{\partial x^2}
    \right)
    + \frac{k_B T_e}{m_i}\frac{\partial}{\partial x}\left\{
        \ln\left[
        \cosh\left(
          \varepsilon b
        \right) \right]
    + \varepsilon b \tanh\left(
          \varepsilon b
        \right)
    \right\} ,
\end{eqnarray}
and

\begin{equation}
  \frac{\partial b}{\partial t} + \frac{\partial}{\partial x}(bV)  = 0 ,
\end{equation}
where we have introduced the Alfv\'en speed $C_A = (B_0^2/\mu_0\rho_0)^{1/2}$ and
the temperature normalized Zeeman energy $\varepsilon = \mu_B B_0/k_BT_e$.

We now normalize our variables as $\bar{t} = \omega_{ci}t$,
$\bar{x} = (\omega_{ci}/C_A)x = (\omega_{pi}/c)x$ (where $\omega_{pi}
= (n_{0i}e^2/\epsilon_0m_i)^{1/2}$ is the ion plasma
frequency), $v = V/C_A$, and $c_s = C_s/C_A$. We then obtain

\begin{eqnarray}
  &&
  \frac{\partial v}{\partial t} 
  + \frac{\partial}{\partial x}\left( \frac{v^2}{2} \right) =
    - \frac{\partial b}{\partial x}
    - c_s^2\frac{\partial}{\partial x}\ln b
  \nonumber \\ &&\qquad\qquad
    + 2\frac{\omega_{pe}^2}{|\omega_{ce}|\omega_C}\frac{\partial}{\partial x}\left(
      \frac{1}{\sqrt{b}}\frac{\partial^2\sqrt{b}}{\partial x^2}
    \right)
    +v_{B}^2 \frac{\partial}{\partial x}\left\{
        \ln\left[
        \cosh\left(
          \varepsilon b
        \right) \right]
          +\varepsilon b \tanh\left(
          \varepsilon b
        \right)
    \right\}
    ,
\label{eq:momfinal}
\end{eqnarray}
with $v_B^2  = k_BT_e/m_iC_A^2 = (1/\varepsilon)(\mu_BB_0/m_iC_A^2)$, and

\begin{equation}
  \frac{\partial b}{\partial t} + \frac{\partial}{\partial x}(bv)  = 0 ,
  \label{continuity}
\end{equation}
where we, for simplicity, drop the bars on the normalized coordinates.

Next, we assume that $v$ and $b$ are functions of $\xi = x -v_0t$,
where $v_0$ is a constant speed (normalized by $C_A$). Then Eq. (\ref{continuity}) can
be integrated as $v = v_0(1 - 1/b)$, where we used the boundary conditions
$b=1$ and $v=0$ at
$|\xi|=\infty$, and Eq. (\ref{eq:momfinal}) can be integrated twice
to obtain 

\begin{equation}
  \left( \frac{dZ}{d\xi}\right)^2 + \Psi(Z) = 0 ,
\end{equation}
where $Z = \sqrt{b}$ and the Sagdeev potential \cite{Sagdeev} for our purposes reads

\begin{equation}
\begin{split}
  &
  \Psi = \frac{|\omega_{ce}|\omega_C}{\omega_{pe}^2}\Bigg\{\frac{v_0^2}{4}\left(
      Z - \frac{1}{Z}
    \right)^2-\frac{1}{4}(Z^2-1)^2-\frac{c_s^2}{2}[Z^2\ln(Z^2)-Z^2+1]
    \\
     &+\frac{v_B^2}{4}
     \bigg(
       Z^2\ln\bigg[\frac{\cosh(\varepsilon Z^2)}{\cosh(\varepsilon)}\bigg]
       -\varepsilon \tanh(\varepsilon) (Z^2-1)
     \bigg)
  \Bigg\} .
  \end{split}
  \label{eq:Sagdeev}
\end{equation}
In deriving (\ref{eq:Sagdeev}) we have used the condition $\Psi(1) = 0$.
In Figs. 1 and 2, we have plotted the Sagdeev potential as well as the profiles of the
corresponding solitary waves for different sets of parameters. The
solitary waves have only sub-Alfv\'enic speeds and are
characterized by a localized depletion of the magnetic field and
density. In Fig. 1, we see that the solitary waves increase their
amplitudes for smaller speeds. In the limit of zero speed, we have rarefactive solitons
with a zero density at its center. The
influence of the electron spin-$1/2$ effect on the solitary waves is displayed in Fig. 2,
where we see that larger values of $\varepsilon$ lead to wider
solitary waves with shallower density and magnetic field depletions.
In order to study the
influence of the spin pressure on the nonlinear dynamics of our system,
we have solved the time-dependent system of equations
(\ref{eq:momfinal}) and (\ref{continuity}) for different values of the
spin pressure parameter $\varepsilon$. 
As an initial condition at $t=0$, we took a magnetic field with a local depletion
in the form Gaussian pulse $b=1-0.5\exp(-x^2/100)$, while the velocity $v$ was set to zero.
For $\varepsilon=5$, we see in the left-hand column of panels in Fig. 3 that
the initial pulse develops into two counter-propagating pairs of 
rarefactive solitary waves, where the smaller pulse in the pair propagates
with a somewhat larger speed, $\sim 0.75\,C_A$, than the larger one that propagates
with a speed of $\sim 0.65\,C_A$. For a larger
value $\varepsilon=10$, displayed in the right-hand panels of Fig. 3, 
the pulse develops into two counter-propagating pulses that propagate with
somewhat lower speed, $\sim 0.4\,C_A$, and they are wider and of smaller 
amplitude than the large-amplitude pulses for $\varepsilon=5$.   
All pulses are rarefactive and are propagating with sub-Alfv\'enic speed, in agreement
with our analysis in Figs. 1 and 2.

\begin{figure}
\centering
\includegraphics[width=10cm]{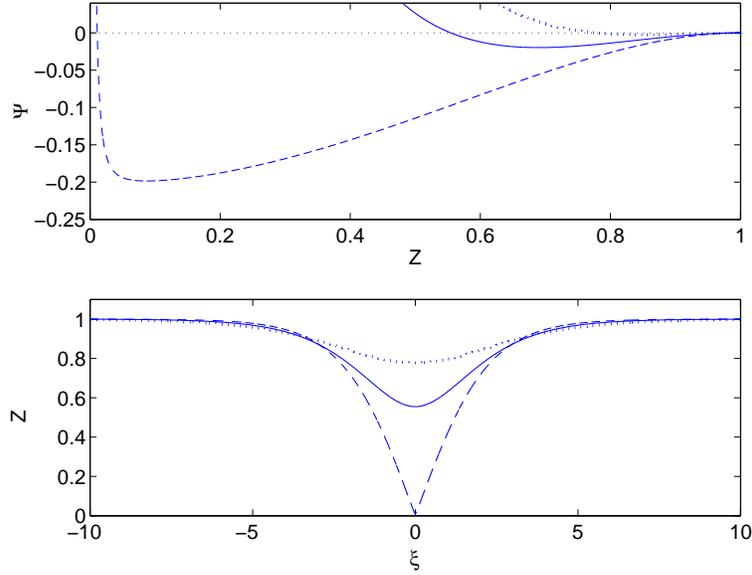}
\caption{The Sagdeev potential $\Psi(Z)$ (upper panel) and the
profile of the solitary wave $Z(\xi)$ (lower panel), for $v_0=0.01$
(dashed lines), $v_0=0.5$ (solid lines) and $v_0=0.7$ (dotted
lines). The other parameters are $\varepsilon=5$, $c_s=0.1$,
$v_B=0.2$ and $|\omega_{ce}|\omega_C/\omega_{pe}^2=1$.}
\end{figure}

\begin{figure}
\centering
\includegraphics[width=10cm]{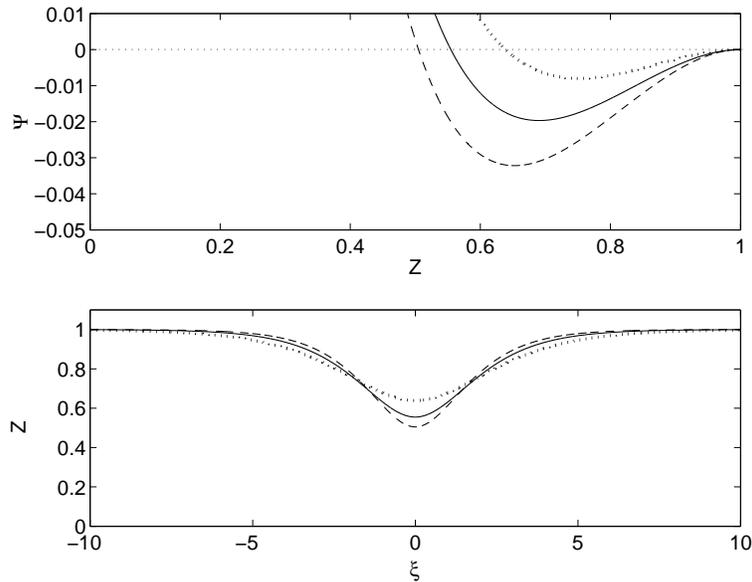}
\caption{The Sagdeev potential $\Psi(Z)$ (upper panel) and the
profile of the solitary wave $Z(\xi)$ (lower panel), for
$\varepsilon=1$ (dashed lines), $\varepsilon=5$ (solid lines) and
$\varepsilon=10$ (dotted lines). The other parameters are $v_0=0.5$,
$c_s=0.1$, $v_B=0.2$ and $|\omega_{ce}|\omega_C/\omega_{pe}^2=1$.}
\end{figure}

\begin{figure}
\centering
\includegraphics[width=10cm]{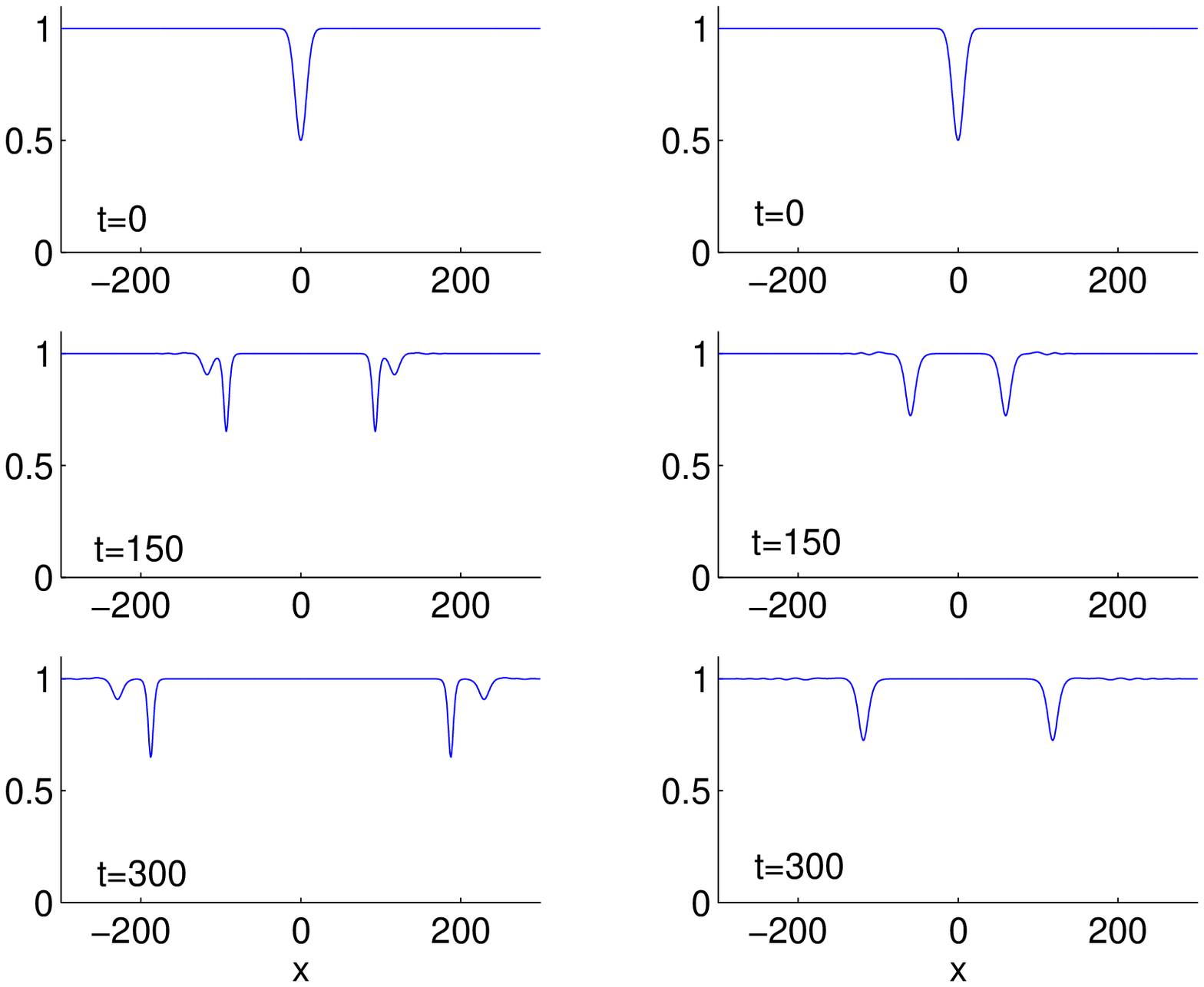}
\caption{The time-dependent dynamics of the normalized magnetic
field magnetic field $b$, for $\varepsilon=5$ (left column) and
$\varepsilon=10$ (right column). The other parameters are 
$c_s=0.1$, $v_B=0.2$ and $|\omega_{ce}|\omega_C/\omega_{pe}^2=1$.}
\end{figure}

%%%%%%%%%%%%%%%%%%%%%%%%%
\section{Summary and Discussion}
%%%%%%%%%%%%%%%%%%%%%%%%%

In the numerical examples of the previous section, the normalized Zeeman energy $\varepsilon$
played a crucial role. In particular, the spin contribution to the soliton dynamics is enhanced when the Zeeman energy is of the order of or greater than one (we note however that other parameters play a role in forming the necessary shape of the Sagdeev potential). Thus, it is natural to investigate what type of parameter values
correspond to $\varepsilon \gtrsim 1$. For astrophysical plasmas, such as in pulsar magnetospheres, we can have $B_0 \lesssim 10^{10}\,\mathrm{T}$ \cite{harding-lai}, implying that the that $\varepsilon \gtrsim 1$ for 
$T_e \lesssim 10^9\,\mathrm{K}$, i.e., not a very severe constraint. However, in such environments, the plasma often has relativistic temperatures and flows, and a relativistic formalism should be used. In the case of Rydberg plasmas \cite{li-etal,fletcher-etal}, where the temperature can go as low as millikelvins, we see that the Zeeman energy is greater than one for external magnetic field $B_0 \gtrsim 10^{-3}\,\mathrm{T}$. Thus, in such ultra-cold laboratory systems, a very weak external magnetic field would make spin effects important for the formation of solitons, and the theory presented here could therefore be checked experimentally.

In conclusion, we have investigated the effects of the quantum Bohm potential and the
electron spin-$1/2$ on the existence of magnetosonic solitary waves in a magnetized
quantum plasma. The solitary waves exist due to a balance between the
nonlinearities and the dispersion induced by the electron quantum
diffraction/tunneling effects associated with the quantum Bohm potential.
The spin introduces an additional negative pressure-like term in the
quantum momentum equation, with the effect that solitary waves become wider and
have shallower density depletions for larger values of the Zeeman energy
$\varepsilon = \mu_BB/k_BT_e$.
We note that the spin term in the Sagdeev potential (13) can dominate the dynamics in the
regime of $C_s^2$, $C_A^2\ll C_A^2v_B^2\varepsilon$. This regime corresponds to a dense 
quantum plasma with an ambient magnetic field, such that
$\omega_{ce}\omega_C\ll\omega_{pe}$ and $k_B(T_e+T_i)\ll \mu_B B_0$. 
Thus, the spin of the electrons collectively modifies the quantum dynamics
of the MHD plasma significantly.

%%%%%%%%%%%%%%%%

\end{document}